\documentclass[final]{ias2}

\usepackage{graphicx} 
\usepackage{multirow}
\usepackage{array} 

\usepackage[colorlinks=false,linktocpage=true]{hyperref}
\usepackage[usenames]{color}
\usepackage[T1]{fontenc}

\def\lesssim{\mathrel{\rlap{\lower4pt\hbox{$\sim$}}
    \raise1pt\hbox{$<$}}} 
\def\gtrsim{\mathrel{\rlap{\lower4pt\hbox{$\sim$}}
    \raise1pt\hbox{$>$}}} 
    
\begin{document}

\markboth{Thermalization and isotropization in heavy-ion collisions}{M. Strickland}

\title{Thermalization and isotropization in heavy-ion collisions}

\author[kent]{Michael Strickland} 
\address[kent]{Physics Department, Kent State University, OH 44242 United States}

\begin{abstract}
I review our current understanding of the processes driving the thermalization and isotropization of the quark gluon plasma (QGP) created in ultrarelativistic heavy ion collisions (URHICs). I begin by discussing the phenomenological evidence in favor of the creation of a thermal but momentum-space anisotropic QGP in URHICs.  I then discuss the degree of isotropization using viscous (dissipative) hydrodynamics, weak-coupling approaches to QGP dynamics, and strong-coupling approaches to QGP dynamics.  Finally, I report on recent progress in the area of real-time non-abelian gauge field simulations and non-abelian Boltzmann-Vlasov-based hard-loop simulations.
\end{abstract}



\maketitle

\section{Introduction}

In this brief review I summarize our current understanding of the thermalization and isotropization of the quark gluon plasma (QGP) created in relativistic heavy ion collisions.  This is still very much an active area of research and, as such, there remain many open questions; however, much has been learned, both on the theoretical and phenomenological fronts since the first $\sqrt{s_{\rm NN}} = 200$ MeV Au-Au data were made available from Au-Au collisions at the Relativistic Heavy Ion Collider (RHIC) at Brookhaven National Lab over a decade ago.  In the interim the heavy ion community has collected a tremendous amount of experimental data and our theoretical understanding, both in terms of our ability to simulate the non-abelian dynamics of the QGP from first principles and to model the QGP based on effective models, has advanced tremendously.  Additionally, with the turn on of the Large Hadron Collider (LHC) at the European Center for Nuclear Research (CERN) in 2008, we now have access to $\sqrt{s_{\rm NN}} = 2.76$ TeV Pb-Pb data which allows us to further push into the QGP part of the phase diagram of quantum chromodynamics (QCD).  Looking to the future, the full energy Pb-Pb runs with $\sqrt{s_{\rm NN}} = 5.5$ TeV will push us even further into the QGP phase.  Despite this progress, there remains an important open theoretical question in the field:  How fast does the QGP thermalize/isotropize and what are the most important processes contributing to this?

There is some lingering confusion concerning the empirical evidence for fast thermalization and isotropization in the QGP.  This confusion stems, in part, from phenomenological fits using ideal hydrodynamics which emerged shortly after the first RHIC data became available.  The heavy-ion community interpreted the ability of ideal hydrodynamical models to describe the $p_T$-dependence of the transverse elliptical flow as solid evidence that the QGP created in heavy ion collisions became isotropic and thermal at approximately 0.5 - 1 fm/c after the initial nuclear impact \cite{Huovinen:2001cy,Hirano:2002ds,Tannenbaum:2006ch,Kolb:2003dz}.  Since the early days of ideal hydrodynamics there was a concerted effort to make hydrodynamical models more realistic by including the effect of shear and bulk viscosities (relaxation times).  This has lead to a proper formulation of relativistic viscous hydrodynamics \cite{Muronga:2001zk,Muronga:2003ta,Muronga:2004sf,Baier:2006um,Romatschke:2007mq,Baier:2007ix,Dusling:2007gi,Luzum:2008cw,Song:2008hj,Heinz:2009xj,El:2009vj,PeraltaRamos:2009kg,PeraltaRamos:2010je,Denicol:2010tr,Denicol:2010xn,Schenke:2010rr,Schenke:2011tv,Shen:2011eg,Bozek:2011wa,Niemi:2011ix,Niemi:2012ry,Bozek:2012qs,Denicol:2012cn} and, recently, anisotropic relativistic viscous hydrodynamics \cite{Martinez:2010sc,Florkowski:2010cf,Ryblewski:2010bs,Martinez:2010sd,Ryblewski:2011aq,Florkowski:2011jg,Martinez:2012tu,Ryblewski:2012rr,Florkowski:2012as,PeraltaRamos:2012xk,Florkowski:2013uqa,Bazow:2013ifa}.  The conclusion one reaches from dissipative hydrodynamics approaches is that the QGP created in ultrarelativistic heavy ion collisions (URHICs) has quite different longitudinal (along the beam line) and transverse pressures, particularly at times $\tau \lesssim 2$ fm/c.

In addition to the progress made in dissipative hydrodynamical modeling of the QGP, there have been significant advances in our understanding of the underlying quantum field theory processes driving the thermalization and (an-)isotropization of the QGP in the weak \cite{Heinz:1985vf,Mrowczynski:1988dz,Pokrovsky:1988bm,Mrowczynski:1993qm,Blaizot:2001nr,Romatschke:2003ms,Arnold:2003rq,Arnold:2004ih,Romatschke:2004jh,Arnold:2004ti,Mrowczynski:2004kv,Rebhan:2004ur,Rebhan:2005re,Romatschke:2005pm,Romatschke:2006nk,Romatschke:2006wg,Rebhan:2008uj,Fukushima:2011nq,Kurkela:2011ti,Kurkela:2011ub,Blaizot:2011xf,Attems:2012js,Kurkela:2012tq,Berges:2012iw,Blaizot:2013lga} and strong coupling  \cite{Chesler:2008hg,Grumiller:2008va,Chesler:2009cy,Albacete:2009ji,Wu:2011yd,Heller:2011ju,Chesler:2011ds,Heller:2012je,vanderSchee:2012qj,Romatschke:2013re,Casalderrey-Solana:2013aba,vanderSchee:2013pia} limits.  The picture emerging from these advances seems to fit nicely into the picture emerging from the aforementioned dissipative hydrodynamics findings, namely that the QGP as created in URHICs possesses large momentum-space anisotropies in the local rest frame, and is particularly anisotropic at early times after the initial nuclear impact.  On the separate issue of thermalization, there is evidence from simulations of weak-coupling non-abelian dynamics that one can achieve rapid apparent longitudinal thermalization of the QGP due to the chromo-Weibel instability~\cite{Attems:2012js} (see also the early time spectra reported in Ref.~\cite{Fukushima:2011nq}); however, there is evidence that this is transient with power-law scaling associated with turbulence emerging on asymptotically long time scales (1000's of fm/c) \cite{Fukushima:2011nq,Berges:2013eia,Berges:2013fga}.  On the strong coupling front, practitioners are now able to use numerical GR to describe the formation of an extra-dimensional black hole (or more accurately an apparent horizon), which is the criterium for QGP thermalization in the AdS/CFT framework.  In an expanding background corresponding to the (approximately) boost-invariant Bjorken-like expansion of the QGP, these studies find thermalization times that are less than 1 fm/c, however, the state which emerges is momentum-space anisotropic even in the infinite 't Hooft coupling limit.

Before proceeding to a more detailed discussion of the (an-)isotropization and thermalization of the QGP, let me point out that URHICs are very much a data-driven field.  Viscous (dissipative) hydrodynamical models are able to describe the collective (elliptic, triangular, etc.) flow of the QGP produced at RHIC and LHC, both in terms of event-averaged observables and their underlying probability distributions, with a surprising level of accuracy.  Since viscous (dissipative) hydrodynamics implies the existence of momentum-space anisotropies in the QGP, one must now conclude, based on empirical evidence, that the QGP might be thermal but strongly anisotropic in momentum-space, implying that the QGP has two temperatures, a transverse one and a longitudinal one.  The existence of such anisotropies must now be taken seriously if one is to treat the phenomenology of the QGP self-consistently.  This means, in practice, that one has to fold into the calculation of various processes, e.g. photon production, dilepton production, heavy quarkonium suppression, jet suppression, etc. the momentum-space anisotropy of the underlying one-particle parton distribution functions.  There have been some initial work along these lines \cite{Romatschke:2003vc,Romatschke:2004au,Schenke:2006yp,Mauricio:2007vz,Martinez:2008di,Dumitru:2007rp,Dumitru:2007hy,Bhattacharya:2008up,Bhattacharya:2008mv,Dumitru:2009ni,Burnier:2009yu,Dumitru:2009fy,Philipsen:2009wg,Bhattacharya:2009sb,Bhattacharya:2010sq,Margotta:2011ta,Strickland:2011mw,Strickland:2011aa,Mandal:2011xn,Mandal:2011jx,Strickland:2012cq,Florkowski:2012ax,Mandala:2013xma} (see also e.g. \cite{Dusling:2008xj,Dusling:2008nt,Dion:2011pp,Shen:2013cca,Shen:2013vja,Vujanovic:2013jpa} for recent progress along these lines using 2nd-order viscous hydrodynamics), but there is much work left to do.  In the process, one may find observables that are sensitive to the level of momentum-space anisotropy in the QGP, thereby allowing us to have independent confirmation of their existence outside of the realm of viscous (dissipative) hydrodynamics.

\section{Momentum-space Anisotropies in the QGP}

As discussed above, many disparate approaches to QGP dynamics consistently find that the QGP, as created in URHICs, possesses large local rest frame momentum-space an-isotropies in the $p_T$-$p_L$ plane due to the initially rapid longitudinal expansion of the matter.  As the first indication of this, let's consider relativistic viscous hydrodynamics for a system that is transversely homogenous and boost invariant in the longitudinal direction, aka 0+1d dynamics.  In this case, first-order Navier Stokes (NS) viscous hydrodynamics predicts that the shear correction to the ideal pressures is diagonal with space-like components $\pi^{zz} = - 4\eta/3\tau = -2\pi^{xx} = - 2\pi^{yy}$, where $\eta$ is the shear viscosity and $\tau$ is the proper time.  In viscous hydrodynamics, the longitudinal pressure is given by ${\cal P}_L = P_{\rm eq} + \pi^{zz}$ and the transverse pressure by ${\cal P}_T =  P_{\rm eq} + \pi^{xx}$.  Assuming an ideal equation of state (EoS), the ratio of the longitudinal pressure over the transverse pressure from first order viscous hydrodynamics can be expressed as
\begin{equation}
\left(\frac{{\cal P}_L}{{\cal P}_T}\right)_{\rm NS} = \frac{ 3 \tau T - 16\bar\eta }{ 3 \tau T + 8\bar\eta } \, ,
\label{eq:aniso}
\end{equation}
where $\bar\eta \equiv \eta/{\cal S}$ with ${\cal S}$ being the entropy density.  Assuming RHIC-like initial conditions with $T_0 = 400$ MeV at $\tau_0 = 0.5$ fm/c and taking the conjectured lower bound $\bar\eta = 1/4\pi$ \cite{Policastro:2001yc}, one finds $\left({\cal P}_L/{\cal P}_T\right)_{\rm NS} \simeq 0.5$.  For LHC-like initial conditions with $T_0 = 600$ MeV at $\tau_0 = 0.25$ fm/c and once again taking $\bar\eta = 1/4\pi$ one finds $\left({\cal P}_L/{\cal P}_T\right)_{\rm NS} \simeq 0.35$.  This means that even in the best case scenario of $\bar\eta = 1/4\pi$, viscous hydrodynamics itself predicts rather sizable momentum-space anisotropies.  For larger values of $\bar\eta$, one obtains even larger momentum-space anisotropies.  In addition, one can see from Eq.~(\ref{eq:aniso}) that, at fixed initial proper time, the level of momentum-space anisotropy increases as one lowers the temperature.  This means, in practice, that as one moves away from the center of the nuclear overlap region towards the transverse edge, the level of momentum-space anisotropy increases.

\begin{figure}[t]
\begin{center}
\includegraphics[width=0.49\textwidth]{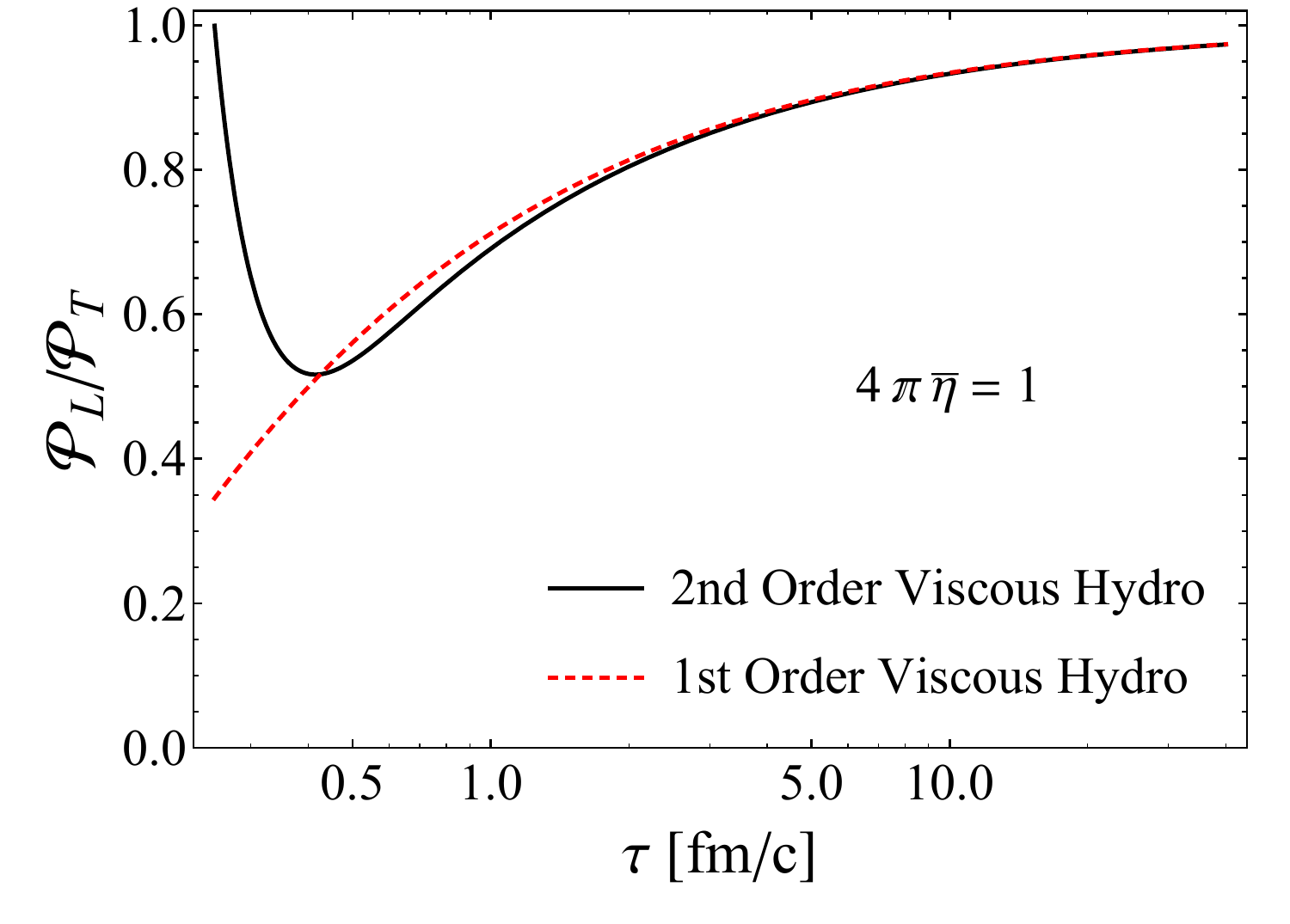}
\includegraphics[width=0.49\textwidth]{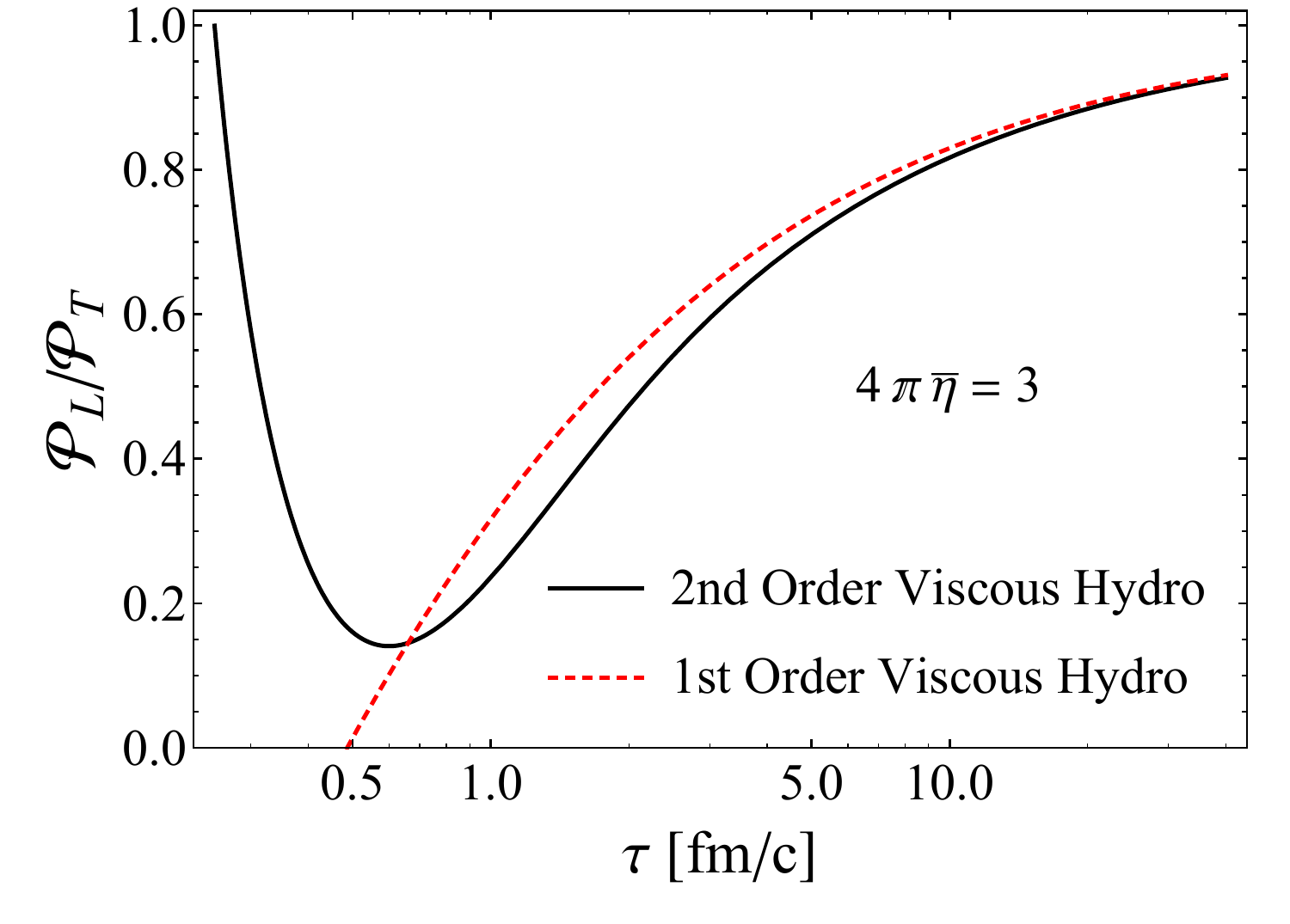}
\end{center}
\vspace{-7mm}
\caption{Pressure anisotropy as a function of proper time assuming an initially isotropic system with $T_0 = 600$ MeV at $\tau_0 =$ 0.25 fm/c for $4\pi\bar\eta =$ 1 (left) and 3 (right).  Solid black line is the solution of the second order coupled differential equations and the red dashed line is the first-order ``Navier-Stokes'' solution.}
\label{fig:nscomp}
\end{figure}

Of course, since first-order relativistic viscous hydrodynamics is acausal, the analysis above is not the full story.  It does, however, provide important intuitive guidance since the causal second-order version of the theory has the first-order solution as an attractive ``fixed point'' of the dynamics.  Because of this, one expects large momentum-space anisotropies to emerge within a few multiples of the shear relaxation time $\tau_\pi$.  In the strong coupling limit of ${\cal N}=4$ SYM one finds $\tau_\pi = (2 - \log 2)/2 \pi T$ \cite{Baier:2007ix,Bhattacharyya:2008jc} which gives $\tau_\pi \sim 0.1$ fm/c and $\tau_\pi \sim 0.07$ fm/c for the RHIC- and LHC-like initial conditions stated above.\footnote{A similar time scale emerges within the kinetic theory framework.}   To demonstrate this quantitatively, in Fig.~\ref{fig:nscomp} I plot the solution of the second order Israel-Stewart 0+1d viscous hydrodynamical equations assuming an isotropic initial condition and the NS solution together.  In the left panel I assumed $4\pi\bar\eta = 1$ and in the right panel I assumed $4\pi\bar\eta = 3$ ($\bar\eta \simeq 0.24$) with $\tau_\pi = 2 (2 - \log 2) \bar\eta/T$ in both cases.  As can be seen from this figure, even if one starts with an isotropic initial condition, within a few multiples of the shear relaxation time one approaches the NS solution, overshoots it, and then approaches it from below.  The value of $\bar\eta$ in the right panel is approximately the same as that extracted from recent fits to LHC collective flow data \cite{Gale:2012rq}.  I note that if one further increases $\bar\eta$, then one finds negative longitudinal pressures in second-order viscous hydrodynamics as well.  

\begin{figure}[t]
\begin{center}
\includegraphics[width=0.44\textwidth]{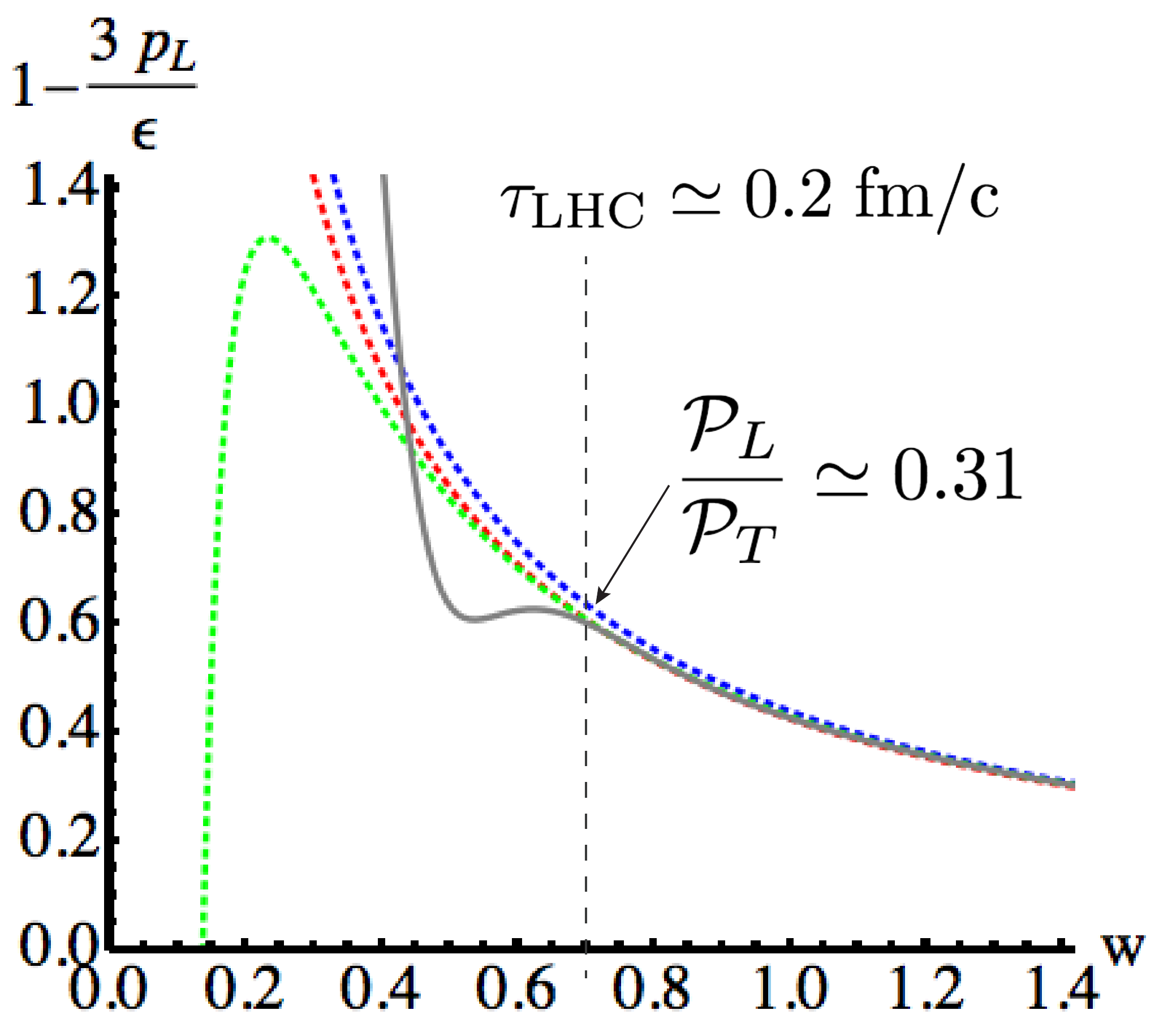}
\includegraphics[width=0.54\textwidth]{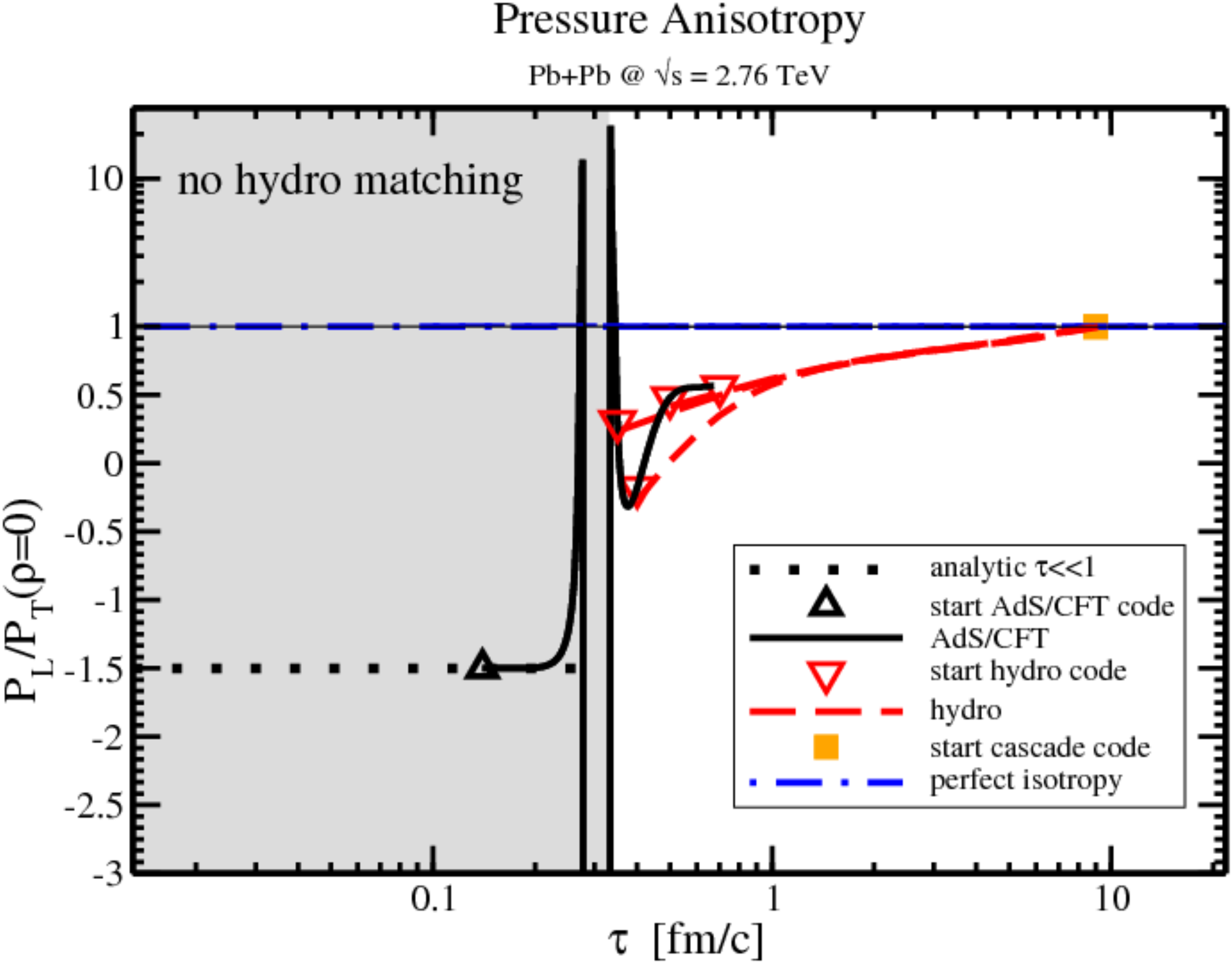}
\end{center}
\vspace{-4mm}
\caption{Pressure anisotropy as a function of proper time from two different AdS/CFT-based calculations.  Left panel shows results from Heller et al \cite{Heller:2011ju} and right panel shows results from van der Schee et al \cite{vanderSchee:2013pia}.  The figure in left panel has been adapted to add labels.}
\label{fig:adsaniso}
\end{figure}

Based on the preceding discussion one learns the value of $\bar\eta$ extracted from LHC data \cite{Gale:2012rq} implies that the system may be highly momentum-space anisotropic with the momentum-space anisotropies persisting throughout the evolution of the QGP.  However, before drawing conclusions based solely on the relativistic viscous hydrodynamics, we can ask the corresponding question within the context of the AdS/CFT framework.  Several groups have been working on methods to address the question of early-time dynamics within the context of the AdS/CFT framework.  Here I focus on the work of two groups:  Heller et al. \cite{Heller:2011ju} and van der Schee et al. \cite{vanderSchee:2013pia} who both simulated the dynamics of an expanding QGP using numerical general relativity (GR).  In the work of Heller et al. they simulated the early time dynamics of a 0+1d system by numerically solving the GR equations in the bulk.  In the work of van der Schee et al. \cite{vanderSchee:2013pia} they performed similar numerical GR evolution but in the case of a 1+1d radially symmetric system including transverse expansion.  

In the left panel of Fig.~\ref{fig:adsaniso} I show the Heller et al. result for the pressure anisotropy expressed as $1 - 3{\cal P}_L/{\cal E}$.  In the paper from which this figure is adapted, the authors found that the evolution begins to be well-approximated by viscous hydrodynamics after a ``time'' of $w = T_{\rm eff} \tau \sim 0.63$ which, upon conversion to physical units using LHC initial conditions, corresponds to $\tau \simeq 0.2$ fm/c (indicated by a vertical dashed line in the figure).  The red, green, and blue dashed lines correspond to first, second, and third order viscous hydrodynamics results and the grey solid line to a typical evolution within their numerical GR approach.  As can be seen from this figure, their results are consistent with the estimates for the initial pressure anisotropy obtained previously.  In addition, we see that the pressure anisotropy persists, decaying as an inverse power of the proper time.  Since their result was obtained in the context of the strong coupling limit for which $4\pi\bar\eta=1$, the pressure anisotropy obtained should be considered an upper bound.  In the right panel of Fig.~\ref{fig:adsaniso} I show the result of 
van der Schee et al.  In this figure, the left shaded region covers the time over which they performed a numerical GR solution which was then matched to viscous hydrodynamical evolution at the border between the grey and white regions.  As can be seen from this figure, even when including radial expansion, one obtains sizable momentum space anisotropies which are consistent in magnitude with the 0+1d results.  Once again the authors assumed $4\pi\bar\eta=1$, so the pressure anisotropy obtained should be considered an upper bound.

Having covered the degree of momentum-space anisotropy predicted by viscous hydrodynamics and the AdS/CFT approach, I would now like to briefly discuss the pressure anisotropies expected within the Color Glass Condensate (CGC) \cite{McLerran:1993ni,McLerran:1993ka,Iancu:2003xm} framework and weakly-coupled gauge field theory in general.  In the CGC framework, the fields are boost-invariant to first approximation.  As a result, the leading order prediction is that longitudinal pressure is zero.\footnote{At $\tau = 0^+$, the longitudinal pressure is negative due to coherent field effects; however, within a few fractions of a fm/c it becomes positive and at leading order goes to zero rapidly.}  Including finite energy corrections results in a very small longitudinal pressure.  Currently, it is believed that the primary driving force for restoring isotropy in the gauge field sector are plasma instabilities such as the chromo-Weibel instability \cite{Strickland:2007fm}; however, so far practitioners have found that, even taking into account the unstable gauge field dynamics, the timescale for isotropization of the system is very long \cite{Romatschke:2006nk,Berges:2007re}.  The recent work of Epelbaum and Gelis \cite{Gelis:2013rba} has included resummation of next-to-leading order (NLO) quantum loop corrections to initial CGC fluctuations, and simulations in this framework find early-time pressure anisotropies on the order of $0.01\,$-$\,0.5$, depending on the assumed value of the strong coupling constant $g_s = 0.1\,$-$\,0.4$.  In the context of hard-loop simulations of chromo-Weibel instability evolution, one finds rapid thermalization of the plasma in the sense that a Boltzmann distribution of gluon modes is established within $\sim$ 1 fm/c (see below); however, large pressure anisotropies persist for at least $5\,$-$\,6$ fm/c \cite{Attems:2012js}.

\section{QGP thermalization via collective instabilities}

There are many processes at play in a non-equilibrium non-Abelian plasma.  These include elastic scattering, inelastic scattering, and collective instabilities.  In the seminal paper of Baier, Mueller, Schiff, and Son (BMSS) \cite{Baier:2000sb} the first two of these were included in a self-consistent calculation of the thermalization and isotropization time of the QGP with the authors of this paper finding that $\tau_{\rm therm} \sim \alpha_s^{-13/5} Q_s^{-1}$ where the constant of proportionality was argued to be approximately 1 \cite{Baier:2000sb}.  Plugging in values of $Q_s$ appropriate for RHIC and LHC energies, 1.4 GeV and 2 GeV, respectively, and boldly extrapolating to $\alpha_s = 0.3$, one finds $\tau_{\rm therm} \sim $ 3.2 fm/c and 2.3 fm/c at RHIC and LHC energies, respectively.  Firstly, we note that according to BMSS, this is the time scale for full isotropization and thermalization of the plasma and so should be taken as an upper limit for the thermalization time since isotropization is much harder (if not impossible) to achieve.  That being said, as mentioned above, this estimate does not include the effect of plasma instabilities and it is natural to ask how do instabilities effect the thermalization and isotropization of the QGP.

Before proceeding let me note that in the context of heavy ion collisions the central question which has to be addressed is that of the impact of longitudinal dynamics on the QGP.  In fact, in the seminal paper of Krasnitz, Nara, and Venugopalan \cite{Krasnitz:2001qu} the authors demonstrated that, within the CGC framework, there is transverse thermalization of gauge fields at times on the order of $Q_s^{-1}$ (few fractions of a fm/c).  They showed that, in the forward light cone, the Coulomb gauge-fixed spectrum of the transverse degrees of freedom was very well described by a 
Bose-Einstein distribution at low momenta and a logarithm-corrected power-law at high-momenta.  This finding implies that the system is already approximately transversally thermal at very early times due to strong gauge field self-interactions.  For this reason, the key questions in URHIC thermalization concern the "longitudinal" thermalization and, more generally, isotropization of a QGP which is (at least approximately) boost-invariant and expanding anisotropically.

The plasma instability which plays the most important role in the isotropization and thermalization of the QGP is the so-called chromo-Weibel instability.\footnote{This instability is named in reference to the analogous Weibel instability which exists in Abelian electromagnetic plasmas~\cite{Weibel:1959}.}  In the asymptotically weak-coupling limit, this instability is present whenever the QGP possesses a certain degree of momentum-space anisotropy.  For a given momentum-space anisotropy, measured by an anisotropy parameter $\xi = \frac{1}{2} \langle p_T^2 \rangle/\langle p_L^2 \rangle - 1$, one finds that a band of soft modes with $k \lesssim g T_\perp$, where $T_\perp$ is the transverse temperature of the system, is initially unstable to filamentation-induced exponential growth of transverse chromomagnetic and chromoelectric fields (primary unstable modes).  Due to non-Abelian interactions, these primary unstable modes rapidly generate longitudinal chromomagnetic and chromoelectric fields which grow at twice the rate as the initially-induced transverse fields \cite{Mrowczynski:2000ed,Romatschke:2003ms,Romatschke:2004jh,Rebhan:2004ur,Schenke:2006fz}.  The growth rate of these unstable modes is parametrically $\Gamma_{\rm instability} \sim g T_\perp$.  Comparing this to other rates, namely the rate for elastic scattering $\Gamma_{\rm elastic} \sim g^4 T_\perp$ and the rate for inelastic scattering and color rotation $\Gamma_{\rm inelastic,color} \sim g^2 T_\perp$, we immediately see that the rate for unstable mode growth exceeds all other relevant processes in the limit of asymptotically small couplings.  As a result, in the weak-coupling limit the dynamics of an anisotropic QGP is dominated by the growth of unstable chromo-Weibel modes.  The investigation of the evolution of soft (gauge) fields subject to dynamical instabilities such as the chromo-Weibel instability \cite{Heinz:1985vf,Mrowczynski:1988dz,Pokrovsky:1988bm,Mrowczynski:1993qm,Blaizot:2001nr,Romatschke:2003ms,Arnold:2003rq,Arnold:2004ih,Romatschke:2004jh,Arnold:2004ti,Mrowczynski:2004kv,Rebhan:2004ur,Rebhan:2005re} is an active area of research.  Field dynamics in an expanding background have been recently investigated using classical Yang-Mills simulations \cite{Romatschke:2005pm,Romatschke:2006nk,Fukushima:2011nq,Fukushima:2011ca,Berges:2012iw,Gelis:2013rba,Berges:2013fga,Fukushima:2013dma}, analytically in the high energy limit \cite{Kurkela:2011ti,Kurkela:2011ub}, within scalar $\phi^4$ theory subject to parametric resonance instabilities \cite{Dusling:2012ig}, and SU(2) Vlasov-Yang-Mills \cite{Romatschke:2006wg,Rebhan:2008uj,Attems:2012js} including longitudinal expansion.  There have also been developments in the area of chromohydrodynamics approaches which also show the presence of the (chromo-)Weibel instability \cite{Basu:2002,Manuel:2006hg,Calzetta:2013nqa}.

The BMMS parametric relation has recently been revisited by Kurkela and Moore (KM) \cite{Kurkela:2011ti,Kurkela:2011ub} to include the effect of the chromo-Weibel instabilities (among many other possibilities which were extensively considered in these papers).  Their conclusion in \cite{Kurkela:2011ti} was that, when putting all of the pieces together, the parametric estimate of the thermalization time of the QGP changes to $\tau_{\rm therm} \sim \alpha_s^{-5/2} Q_s^{-1}$; however, they did not provide an estimate for the constant of proportionality.  In terms of the exponent of $\alpha_s$, one finds in the BMSS scenario 13/5 = 2.6 and in the KM scenario one finds instead 5/2 = 2.5; however, the uncertainty in the constant of proportionality remains, which could significantly change things.  Assuming that this constant is order 1, one finds that in the weak-coupling limit plasma instabilities accelerate the thermalization of the QGP, but not dramatically.  Once again, however, associating one number with both thermalization and isotropization is probably too limiting since evidence to date indicates that the plasma may become thermal on a shorter time scale than it becomes isotropic in momentum space (at least for the soft momenta that viscous hydrodynamical modeling can reliably describe).  

In the weak-field regime with a fixed momentum-space anisotropy, 
the chromo-Weibel instability initially causes exponential growth of transverse
chromomagnetic and chromoelectric fields; however, due to non-Abelian interaction between the fields, 
exponentially growing longitudinal chromomagnetic and
chromoelectric fields are induced which grow at twice the rate of the transverse field configurations.
Eventually, all components of the unstable gauge-field configurations become of equal magnitude.
As a result, one finds strong gauge field self-interaction at late times and numerical simulations are necessary
in order to have a firm quantitative understanding of the late-time behavior of the system
\cite{Arnold:2003rq,Rebhan:2004ur,Arnold:2005vb,Rebhan:2005re,Arnold:2005ef,Arnold:2005qs,Romatschke:2006nk,Fukushima:2006ax,Bodeker:2007fw,Arnold:2007cg,Berges:2007re,Berges:2008mr,Berges:2008zt,Berges:2009bx,Ipp:2010uy,Dusling:2011rz,Berges:2012iw}.
In order to understand the precise role played by the chromo-Weibel instability in ultrarelativistic heavy ion
collisions, one must include the effect of the strong longitudinal expansion of the matter.  For
the first few fm/c of the QGP's lifetime, the longitudinal expansion dominates the transverse
expansion.  Therefore, to good
approximation, one can understand the early time dynamics of the quark gluon plasma by considering only
longitudinal expansion.  The first study to look at the effect of longitudinal expansion was done in the
context of pure Yang-Mills dynamics initialized with color-glass-condensate initial conditions onto which
small-amplitude rapidity fluctuations were added \cite{Romatschke:2006nk}.  
The initial small-amplitude fluctuations result from quantum corrections to the classical dynamics 
\cite{Fukushima:2006ax,Fukushima:2011nq,Dusling:2011rz}.  Numerical studies have shown that adding spatial-rapidity 
fluctuations results in growth of chromomagnetic and chromoelectric fields with amplitudes
$\sim \exp(2 m_D^0\sqrt{\tau/Q_s})$ where $m_D^0$ is the initial Debye screening mass and $\tau$
is the proper time.  This growth with $\exp(\sqrt{\tau})$ was predicted by Arnold et al. based on 
the fact that longitudinal expansion dilutes the density~\cite{Arnold:2003rq}.

In a recent study within the hard-loop framework \cite{Attems:2012js} the authors assumed that the background particles are longitudinally free streaming and, as a result, the background (hard) particles possess a local rest frame momentum-space anisotropy which increases quadratically in proper-time. 
Given an isotropic distribution $f_{\rm iso}$, the corresponding longitudinal free-streaming one-particle distribution function can be straightforwardly constructed.  Following \cite{Romatschke:2006wg} one can obtain the dynamical equations obeyed by color perturbations $\delta\!f^a$ of a color-neutral longitudinally free-streaming momenta distribution $f_0$
$
V\cdot  D\, \delta\!f^a\big|_{p^\mu}=g  V^\mu
F_{\mu\nu}^a  \partial_{(p)}^\nu f_0(\mathbf
p_\perp, p_\eta) .
\label{Vlasov}
$
This equation must be solved simultaneously together with the non-Abelian Yang-Mills equations which couple the color-charge fluctuations back to the gauge fields via the induced color-currents $j^\nu_a$
\begin{equation}
D_\mu  F^{\mu \nu}_a  = j^\nu_a = 
g\, t_R \int{\frac{d^3p}{ (2\pi)^3}} \frac{p^\mu}{2 p^0} \delta\!f_a(\mathbf p,\mathbf x,t) \, , 
\label{Maxwell}
\end{equation}
where $D_\alpha=\partial_\alpha-ig[ A_\alpha,\cdot]$ is the gauge covariant derivative, $F_{\alpha\beta}=\partial_\alpha  A_\beta-\partial_\beta  A_\alpha -ig[ A_\alpha, A_\beta]$ is the gluon field strength tensor, and $g$ is the strong coupling.
The above equations are then transformed to comoving coordinates with the metric $ds^2=d\tau^2-d\mathbf x_\perp^2-\tau^2 d\eta^2$.

\begin{figure}[t]
\centerline{%
\includegraphics[width=0.5\textwidth]{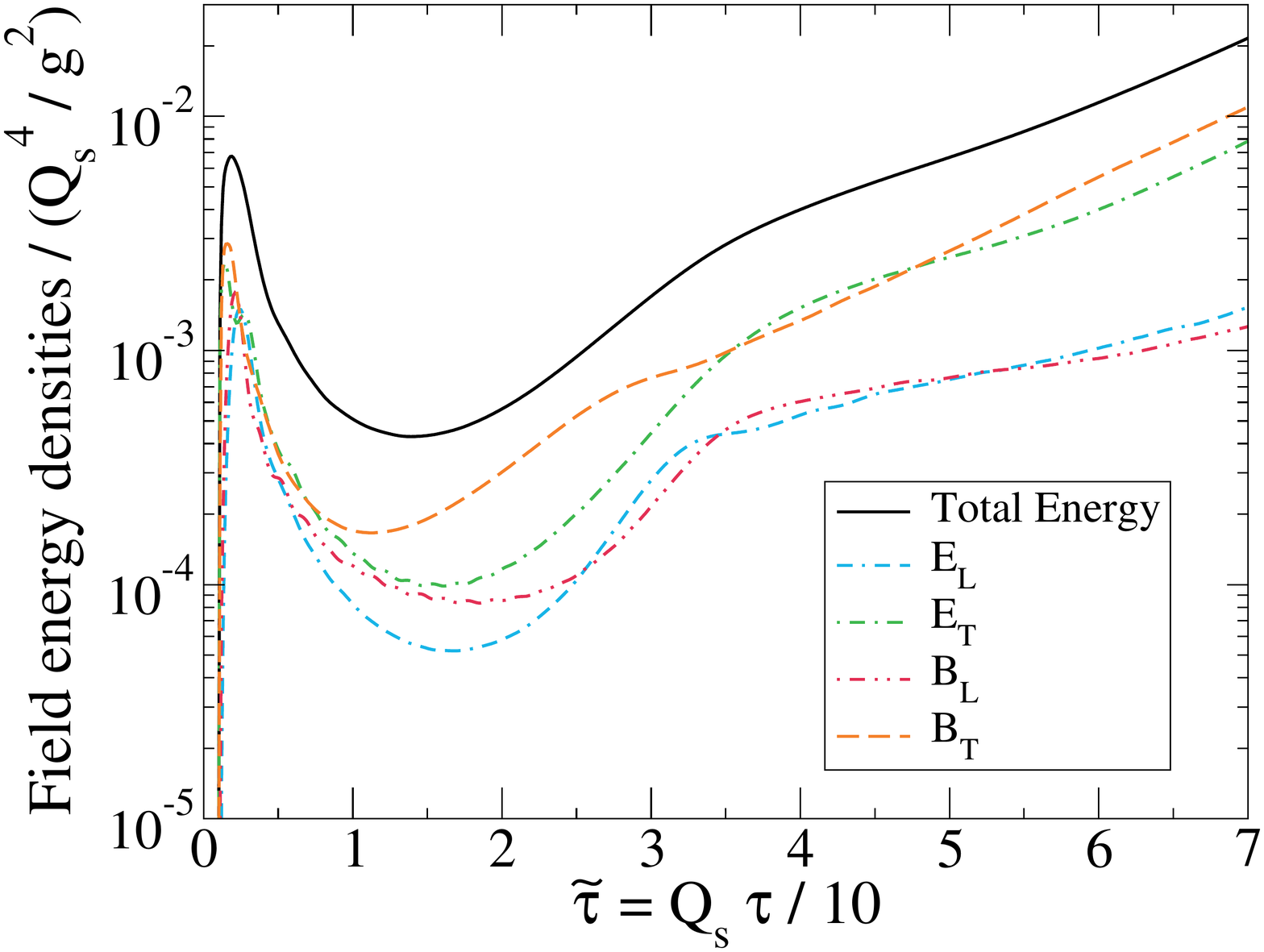}
\includegraphics[width=0.5\textwidth]{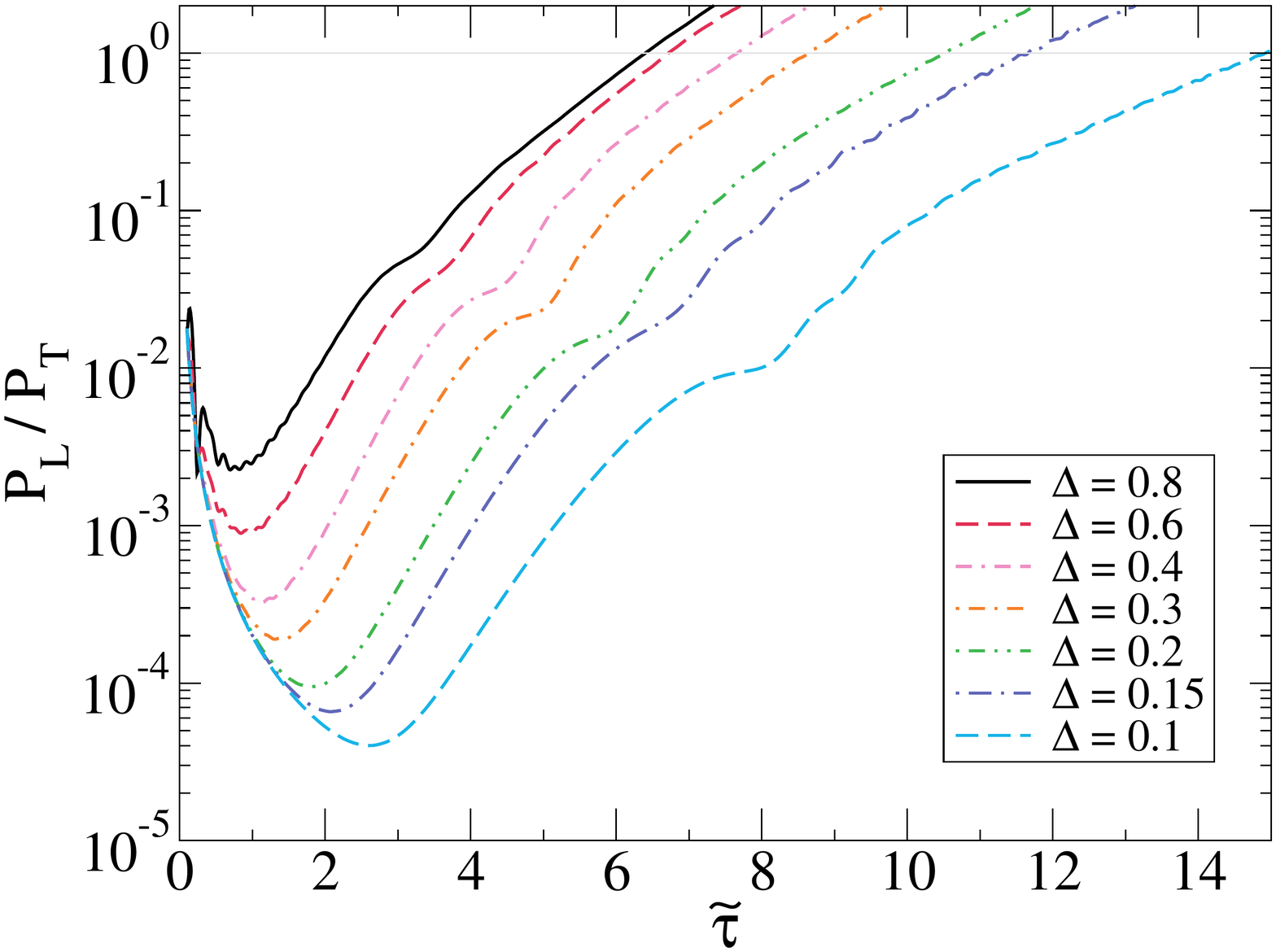}
}
\caption{On the left I plot the various components of the chromofield energy density as a
function of proper time.  On the right I plot the total (field plus particle) longitudinal
over transverse pressure as a function of proper time.}
\label{fig:egrun}
\end{figure}

The resulting dynamical equations are numerically solved in temporal axial gauge on a spatial lattice.  In order to maintain gauge invariance with respect to three-dimensional gauge transformations, the spatially-discretized fields are represented by plaquette variables and evolved along with the conjugate momentum using a leap-frog algorithm.  The fluctuation-induced currents are represented by auxiliary fields which are discretized in space and also on a cylindrical velocity-surface spanned by azimuthal velocity and rapidity.  As a result, the simulations are effectively five-dimensional and are, therefore, computationally intensive.  For details concerning the numerical implementation see Ref.~\cite{Attems:2012js}.
For the initial conditions Ref.~\cite{Attems:2012js} seeded current fluctuations of amplitude $\Delta$ which had a  UV spectral cutoff.  In Fig.~\ref{fig:egrun} (left) I show 
the various components of the chromofield energy density as a function of rescaled proper time $\tilde\tau$.  For LHC 
and RHIC initial energy densities, one unit in $\tilde\tau$ corresponds to approximately 1 fm/c and 1.4 fm/c, 
respectively.  For this figure an initial fluctuation amplitude of $\Delta = 0.8$ was chosen.  

As can be seen from this figure, after approximately 1 fm/c we begin to see rapid growth of the transverse 
chromomagetic field, followed by the transverse chromoelectric field, and then the longitudinal chromofields.
In Fig.~\ref{fig:egrun} (right) we show the resulting ratio of the total (particle plus field) longitudinal 
pressure divided by the total transverse pressure for various values of $\Delta$.  At early times, 
prior to unstable mode growth, one observes that the longitudinal pressure drops, due to the longitudinal free streaming 
of the hard particle background; however, when the unstable modes have grown significantly, one observes a regeneration of the
longitudinal pressure by the unstable modes.  In addition, one observes that the time at which isotropy is restored is primarily sensitive to
the initial fluctuation amplitude $\Delta$.

\begin{figure}[t]
\centerline{%
\includegraphics[width=\textwidth]{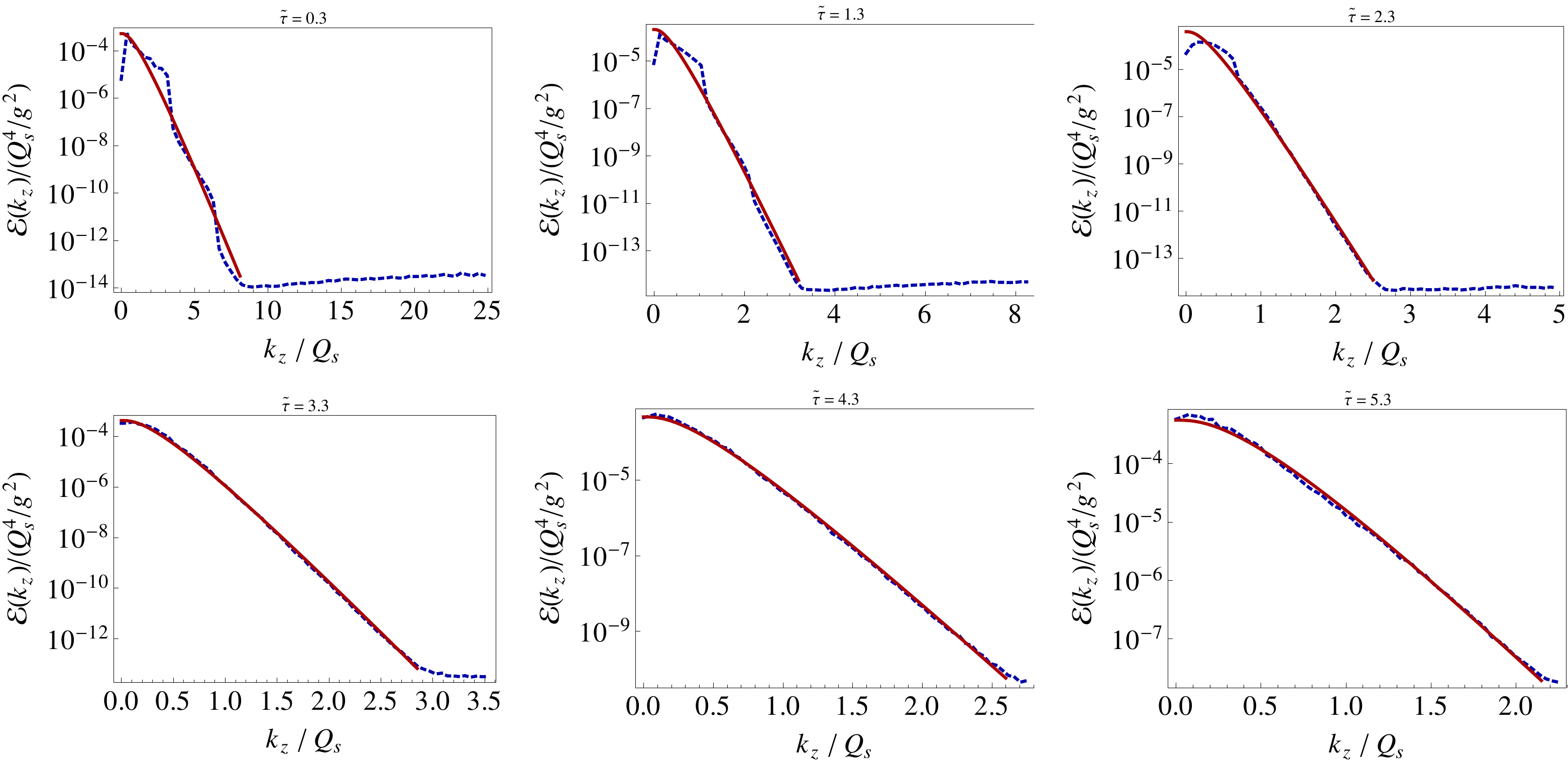}
}
\caption{On the left I plot the longitudinal spectra at various proper times.  On the right I
plot the extracted longitudinal temperature which was obtained by a fit (see text) to the longitudinal
spectra ($\cal E$) or the Fourier transform of the spatial energy density ($\overline{\cal E}$).}
\label{fig:spectra}
\end{figure}

In addition to extracting information about the energy density and pressures of the system as a function
of proper time, one can also extract information from the gauge field spectra.
The longitudinal spectra can be obtained following Ref.\
\cite{Fukushima:2011nq} by first Fourier transforming each field component,
integrating over the transverse wave vectors and
decomposing each according to the longitudinal wave vector $\nu$,
in terms of which the electric and magnetic energy densities
are decomposed into longitudinal energy spectra (see Ref.~\cite{Attems:2012js} for details).
In Fig.~\ref{fig:spectra} (left) I show the extracted longitudinal spectra extracted using the first method averaged over 50 runs.  
The lines shown in panel are fits to a form ${\cal E} \propto \int d k_z \left( k_z^2 + 2 |k_z| T + 2 T^2 \right) \exp\left(-|k_z|/T\right)$
which is obtained by integrating a Boltzmann distribution over transverse momenta.  As can be seen from these panels
this fit function begins to describe the
observed spectra very well at early times corresponding to $\tilde\tau \sim 1$ indicating 
extremely fast longitudinal thermalization of the spectra even though the system is still
highly anisotropic at this moment in time.

\section{Conclusions and Outlook}

In this brief review, I have attempted to discuss recent advances and outstanding questions regarding our
theoretical understanding of the thermalization and isotropization of the QGP.  As pointed out herein,
at this moment in time all signs indicate that the QGP created in URHICs is anisotropic in momentum space with large anisotropies
expected at early times and near the transverse edges of the plasma.  These anisotropies last for multiple fm/c
and, as a result, modern phenomenological approaches should include these anisotropies in e.g. production matrix 
elements, quark energy loss, and quarkonium potentials.\footnote{In the context of viscous hydrodynamics this 
translates into including the viscous corrections to the thermal one-particle distribution function self-consistently
in the hydrodynamic simulation as well as the process under consideration.}  Despite the momentum-space anisotropy
in the local rest frame, there are indications 
from both the weak- and strong-coupling approaches that the system thermalizes in the sense that there is
(at least transiently) a Boltzmann-like distribution of energy or the formation of an apparent horizon in the bulk, respectively.

On the weak-coupling side, calculations and simulations are increasing their scope and associated complexity. The simulations required 
are numerically intensive due to high-dimensionality, in the case of hard-loop codes, and the large lattice sizes and statistical averaging
required in general.  As a result, the time scale for advances in our understanding of weak-coupling dynamics has grown longer 
recently.  On the strong-coupling side, there have been significant advances in our
understanding of strong-coupling thermalization and (an)isotropization of the QGP.  The state of the art calculations
now include azimuthally symmetric transverse expansion for smooth initial conditions and are able to interpolate between
full stopping and boost-invariant Bjorken flow based on the initial condition chosen.  As these simulations become
more realistic and eventually start to include fluctuations in the initial conditions, they too will face some
rather daunting numerical problems, but these are not insurmountable.

\section*{Acknowledgments}

\vspace{-2mm}
This work was supported in part by DOE Grant No.~DE-SC0004104.

\bibliographystyle{pramana}
\bibliography{qgptherm}

\end{document}